\documentclass{article}    % Specifies the document style.
\usepackage{amsmath,amsthm,amsfonts}

\usepackage{graphicx}
\newcommand{\ib}{\int_B}

\newcommand{\ibrd}{\bar{I}^{\dag}}

\newcommand{\sbr}{\bar{S}}

\begin{document}
           % End of preamble and beginning of text.
\pagestyle{empty}

\begin{flushleft}
{\Large \bf Applications of the Nonequilibrium Kubo Formula to the
Detection of Quantum Noise}\\ \vspace{0.3cm}

%(Invited talk to be given by Y. I. at the SPIE Noise Conference, Grand
%Canary, June 2004, accepted for the proceedings.)

\end{flushleft}

\noindent U. Gavish$^1$, {\underline {Y. Imry}}$^2$, B.
Yurke$^{3,4}$\\ \\ 1. LKB, Ecole Normale Superieure, Paris\\ 2.
Condensed Matter Physics Dept., Weizmann Institute, Rehovot \\ 3.
Bell Laboratories,
Lucent Technologies, Murray Hill,  NJ\\ 4. Caltech. Pasadena, Cal. \\

\textbf{Abstract.} The Kubo fluctuation-dissipation theorem
relates the current fluctuations of a system in an equilibrium
state with the linear AC-conductance. This theorem holds also out
of equilibrium provided that the system is in a stationary state
and that the linear conductance is replaced by the (dynamic)
conductance with respect to the non equilibrium state. We provide
a simple proof for that statement and then apply it in two cases.

We first  show that in an excess noise measurement at zero
temperature, in which the impedance matching is maintained while
driving a mesoscopic sample out of equilibrium, it is the {\it
nonsymmetrized} noise power spectrum which is measured, even if
the bare measurement, i.e. without extracting the excess part of
the noise,  obtains the symmetrized noise.

As a  second application we derive a commutation relation for the
two components of fermionic or bosonic currents which holds in
every stationary state
%(and \emph{also} for bosonic currents) and
and which is a generalization of the one valid \emph{only} for
bosonic currents. As is usually the case, such a commutation
relation can be used e.g. to derive Heisenberg uncertainty
relationships among these current components.

\section{Introduction: Definitions, Kubo formula}
\label{sec 1} Consider first a system in an equilibrium state and
its current operator  defined by:
\begin{eqnarray}
 \label{I def} I=\frac{1}{L}\int_{L^3} dx j(x)\end{eqnarray}
 where
\begin{eqnarray}\label{j def} j(x)=\frac{e}{2m}\sum_i (P_{i} \delta(x-x_i)+ \delta(x-x_i)P_{i}).\end{eqnarray}
$e$ is the charge of each of the particles in the system, $P_i$
its momentum and $x_i$ its position. We consider the current in a
cube of volume $L^3.$ For simplicity we take $L^3$ to be a unit
cube, $L=1,$ and write our formulae in one dimension\cite{3D}.

The current fluctuations are often described by the \emph{noise
power spectrum} defined by
\begin{eqnarray}
\label{S def} S(\omega)=\int_{-\infty}^{\infty} dt e^{i\omega t}
\langle I(0)I(t)\rangle.\end{eqnarray}
  $\langle ...\rangle$ denotes averaging with respect to a stationary state:
\begin{eqnarray}
\label{ave def} \langle A \rangle =Tr \rho_0 A \end{eqnarray}
 where $\rho_0$ is a time independent density matrix:
\begin{eqnarray}
\label{ro def} \dot{\rho}_0=0, ~~~~~[H,\rho_0]=0\end{eqnarray}
 and $H$ is the Hamiltonian.

Suppose now that the system is driven out of equilibrium by
applying an AC electric field
\begin{eqnarray}
\label{E def}E(t)=Ee^{i\omega t}.\end{eqnarray}
 For a weak enough $E(t)$ (this regime is called the linear response regime) the resulting current will be of the form
\begin{eqnarray}
 \label{sig def} I(t)=\sigma (\omega) Ee^{i\omega t}\end{eqnarray}
 where $\sigma (\omega)$ is time independent. We define \emph{the conductance} of the system, $G_d(\omega),$ by
\begin{eqnarray}
 \label{Gd def} G_d(\omega)=Re \sigma (\omega). \end{eqnarray}
 In  more general situations, one can perturb the system by various external fields and measure other properties beside
the electrical current. In such cases $G_d(\omega),$ (or
$\sigma(\omega)$) is called \emph{the linear-response
}coefficient. Here we shall focus on the electrical current but
our discussion is extendable to the general case.

Let $S_{eq}(\omega)$ denote $S(\omega)$ of a system  in
equilibrium. In 1956 Kubo \cite{kubo0}-\cite{Kubo2} derived a
fluctuation-dissipation theorem which relates  $G_d(\omega),$ and
$S_{eq}(\omega).$ It is the following:
\begin{eqnarray}
\label{kubo} S_{eq}(-\omega)-S_{eq}(\omega)=2 \hbar\omega
G_d(\omega).\end{eqnarray}
 Justifiably, Kubo called it a fluctuation-dissipation relation since it relates the dissipative properties of the system, $G_d(\omega),$
with its equilibrium fluctuations.  There exists also another
relation of this type which was derived by Callen and Welton
\cite{FDT} in 1951, and which is widely known as \emph{the}
fluctuation-dissipation theorem. It is:
\begin{eqnarray}
 \label{callen welton} \frac{1}{2}(S_{eq}(-\omega)+S_{eq}(\omega))=
2G_d(\omega)(\frac{\hbar\omega}{2}+\frac{\hbar\omega}{ e^{
\hbar\omega /k_{B} T}-1}) \end{eqnarray}
 where $T$ is the temperature.
On one hand, the Callen-Welton relation is valid only for a system
in equilibrium. On the other hand, in his work Kubo stressed that
Eq.\ref{kubo} enables a prediction of a nonequilibrium property
such as the conductance by a calculation of an equilibrium one
\cite{Onsager} (Although he did not rule out generalization to
nonequilibrium). These are probably the two main reasons why it is
often believed that Eq.\ref{kubo} is not valid for nonequilibrium
states. However, Eq.\ref{kubo} \emph{is} valid in any
nonequilibrium state \cite{Landau Lifshitz kubo},\cite{moriond
2001},\cite{gavish thesis} provided that this state is stationary.
That is,
\begin{eqnarray}
\label{kubo noneq} S(-\omega)-S(\omega)=2\hbar\omega
\tilde{G}_d(\omega).\end{eqnarray}
 Here $S(\omega)$ is given by Eq.\ref{S def} at any stationary nonequilibrium state (i.e. with the condition \ref{ro def}).
 $\tilde{G}_d(\omega)$ is the response with respect to a small perturbation which is applied to the system which is already driven out of equilibrium
by another, not necessarily small, perturbation. Like
Eq.\ref{kubo}, Eq.\ref{kubo noneq} holds also for interacting
systems.

For example, consider a mesoscopic system at zero temperature
which is driven out of equilibrium by an external DC field. As  a
result a DC current arises. $S(\omega)$ will then be the
nonsymmetrized shot-noise spectrum related to this current.
Suppose now that an \emph{additional} small "tickling" AC field
$E(t)=E e^{i\omega t}$ is applied on top of the DC one. As a
result also an additional current appears:
\begin{eqnarray}
\label{sig d def}\Delta I(t)\equiv \langle I(t)
\rangle_{E>0}-\langle I \rangle_{E=0}=\tilde{\sigma}(\omega)
Ee^{i\omega t}\end{eqnarray}
 and now
\begin{eqnarray}
 \label{Gd noneq def} \tilde{G}_d(\omega)=Re \tilde{\sigma} (\omega), \end{eqnarray}
 that is, $\tilde{G}_d(\omega)$ will then  be the linear coefficient relating the new field with the new current (it is therefore perhaps
more appropriate to call it the differential AC-conductance to
distinguish it from the one valid when the AC field is applied in
equilibrium). Eq.\ref{kubo noneq} relates the shot-noise spectra
and this differential AC-conductance.

In section \ref{sec 2} we give a simple, self contained,
derivation of Eq.\ref{kubo noneq}. In section \ref{sec 3} it is
shown that noise measurement setup that measures the symmetrized
power spectrum at zero temperature yields the nonsymmetrized one
when used in an excess noise measurement (provided that the
system-setup impedance matching is kept constant while the system
is driven out of equilibrium). This is a direct consequence of
Eq.\ref{kubo noneq}. In section \ref{sec 4}  Eq.\ref{kubo noneq}
is used to generalize the canonical commutation relations valid
for a current in a boson field to the case of a fermionic one,
provided the commutator is replaced by its expectation value in a
stationary state.

\section{Derivation of the nonequilibrium Kubo formula}
\label{sec 2} Eq.\ref{kubo noneq} was obtained in ref.
\cite{moriond 2001} by calculating the net absorption from a
classical
 EM field and using the relation between this dissipation and the conductance. A more mathematical proof is given in Ref.\cite{Landau Lifshitz kubo}.
Here we present a simple and systematic derivation which follows
closely the original one given by Kubo \cite{Kubo1}, except that,
we do not make use of the specific form of the density matrix has
in equilibrium but only assume it to be time independent as in any
stationary state. Consider a system described by the Hamiltonian
\begin{eqnarray}
\label{H0} H_0=\sum_{i=1}^n \frac{P_i^2}{2m}+V \end{eqnarray}
 where $V=V(x_1,..,x_n).$
To describe the application of a small external alternating
electrical field we rewrite it as usual as
\begin{eqnarray}
\label{H} H =\sum_{i=1}^n \frac{(P_i-e
A(x_i,t))^2}{2m}+V(x_1,..x_n),\end{eqnarray}
 (throughout this section we take $c=1$).
The scalar potential does not appear since we are using the
transverse gauge. In this gauge one has
\begin{eqnarray}
\label{maxwell} E(x,t)=-\dot{A}(x,t).\end{eqnarray}
 We write $A(x_i,t)=\int dx A(x,t)\delta(x_i-x)$ (so now $ A(x,t)$ is no longer an operator
since $x_i$ is in the $\delta$-function) and keep only first order
term in $A$. $H$ becomes
\begin{eqnarray}
\label{H lin} H=\sum_{i=1}^n \frac{P_i^2}{2m}+V -\int dx
A(x,t)j(x)\end{eqnarray}
 where $j(x)$ is given by Eq.\ref{j def}.
We now assume that $A(x,t)$ is constant within the cube $L^3,$ and
vanishes outside of it. Adding  another part of $A(x,t)$, simply
results in adding the linear response to it, so this assumption is
not a restrictive one. It is needed only because we are looking
for the conductance which is related to $I(t)$ given by Eq.\ref{I
def}. With the above assumption we have $\int dx A(x,t)j(x)=A(t)
\int_{L^3} j(x)$ and hence (recall $L=1$):
\begin{eqnarray}
\label{H lin final} H=H_0 -A(t) I.\end{eqnarray}
 Since we are looking for a relation for a single frequency $\omega$ we consider the case in which $A(t)$ is of the form:
\begin{eqnarray}
 \label{A om}A(t)= A e^{i\omega t}\end{eqnarray}
 and thus, by Eq.\ref{maxwell},
\begin{eqnarray}
 \label{E om} E(t)= -i\omega A e^{i\omega t}.
\end{eqnarray}
 By Eq.\ref{sig d def} we have\footnote{$I$ appearing in Eq.\ref{d It 1} is the same as in Eq.\ref{I def}, i.e., it is the average of $j(x),$  Eq.\ref{j def}.
In the presence of the vector potential the proper (gauge-invariant) current is given by $j_A(x)=j(x)-\frac{e}{m}\rho(x)A(t)$
where $\rho(x)=\sum_i \delta(x-x_i)$ is the density and thus one should replace $I$ by $I_A\equiv I -\frac{e}{m}QA(t)$
where $Q$ is the total charge. Since the extra (diamagnetic) term $\frac{e}{m}QA(t)$ is linear in $A$ it may affect the linear response.
However, by Eq.\ref{E om}, $\frac{e}{m}QA(t)=i\frac{e}{m\omega}QE(t).$
Because of the $i$ in front of the real coefficient, $\frac{e}{m\omega}Q,$
this term contributes only to the out-of-phase (non-dissipative) part of the current. In other words,
defining (in analogy with Eq.\ref{sig d def}):
$\Delta I_A(t)=\tilde{\sigma}_A(\omega)Ee^{i\omega t},$
one obtains from all the above $Re\tilde{\sigma}_A(\omega)=Re \tilde{\sigma}(\omega).$
Only the real part of $\tilde{\sigma}(\omega),~$ $\tilde{G}_d(\omega),$ appears in Eq.\ref{kubo noneq}, and therefore our use of $I$
instead of $I_A$ is justified.}:
\begin{eqnarray}
\label{d It 1}\Delta I(t)=-\tilde{\sigma}(\omega) i\omega A
e^{i\omega t}.\end{eqnarray}
 and also
\begin{eqnarray}
\label{d It 2}\Delta I(t)=Tr \delta\rho(t) I\end{eqnarray}
 where
\begin{eqnarray}
\label{dl ro def}  \rho(t)-\rho_0=\delta \rho(t) \sim
O(A)\end{eqnarray}
 is the change in the density matrix due to switching on the perturbation.
The equation of motion of the density matrix is
\begin{eqnarray}
\label{ro eom1} \dot{\rho}(t)=-\frac{i}{\hbar}
[H,\rho(t)].\end{eqnarray}
 Recalling that the same equation holds for $\rho_0,$ that $\delta \rho(t) \sim O(A),$ and keeping only first order terms in $A$ we get
\begin{eqnarray}
\label{ro eom2} \delta\dot{\rho}(t)=-\frac{i}{\hbar}
[H_0,\delta\rho(t)]+\frac{i}{\hbar}A(t)[I,\rho_0].\end{eqnarray}
 Eq.\ref{ro eom2} is solved by substituting
$\delta \rho(t)= e^{-\frac{i}{\hbar}H_0t}\alpha
(t)e^{\frac{i}{\hbar}H_0t},$ solving for $\alpha(t),$ expressing
the result in terms of $\delta\rho(t)$ and then using Eq.\ref{ro
def}.  One obtains
\begin{eqnarray}
\label{dl ro solve} \delta\rho(t)=\frac{i}{\hbar}\int_0^{\infty}
d\tau A(t-\tau)[I(-\tau),\rho_0]\end{eqnarray}
 where
\begin{eqnarray}
\label{It def} I(t)=e^{\frac{i}{\hbar}H_0t}I
e^{-\frac{i}{\hbar}H_0t}\end{eqnarray}
 is the Heisenberg current operator of the \emph{unperturbed} system. Inserting this result into Eq.\ref{d It 2} and using $Tr ABC=Tr CAB$
one gets
\begin{eqnarray}
\label{dl I final}\Delta I(t)= \frac{i}{\hbar}\int_0^{\infty}
d\tau A(t-\tau)\langle [I(0),I(-\tau)]\rangle.\end{eqnarray}
 Comparing with Eq.\ref{d It 1} and using Eq.\ref{A om} yields
\begin{eqnarray}
\label{sig d final} \hbar\omega\tilde{\sigma}(\omega)
=-\int_0^{\infty} d\tau
e^{-i\omega\tau}\langle[I(0),I(-\tau)]\rangle.\end{eqnarray}
 Finally, taking the real part of the last equation while using $I^{\dag}=I$ and the fact that in
 a stationary state one has $\langle I(0)I(-\tau)\rangle=\langle I(\tau)I(0)\rangle,$
we obtain
\begin{eqnarray}
\label{kubo noneq final}
S(-\omega)-S(\omega)=\int_{-\infty}^{\infty} d\tau e^{i\omega \tau
} \langle [I(\tau),I(0)]\rangle = 2\hbar\omega
\tilde{G}_d(\omega)\end{eqnarray}
 as in Eq.\ref{kubo noneq}.

\section{First application. Excess noise measurement}
\label{sec 3} Consider a mesoscopic system at zero temperature
coupled to a detection setup (also at zero temperature) which is
designed in such a way that it will measure (as is very often {\it
assumed}) the symmetrized noise spectrum:
\begin{eqnarray}
\label{S sym def} S_{m}(\omega)=S_{sym}(\omega)=\frac{1}{2}
(S(\omega)+S(-\omega)).\end{eqnarray}
 $S_m(\omega)$ stands for the measured spectrum.
Such a setup may resemble, for example, the one used in
\cite{koch}. In \cite{Gavish Delft} it was shown that for a very
broad class of setups, if one subtracts the noise measured at
equilibrium from the nonequilibrium one the resulting spectrum
will be given by Eq.\ref{S def}, i.e. it will be nonsymmetrized\footnote{This was shown also for setups that include
 an amplification stage (as in \cite{koch}), which is usually the one determining the measured
quantity (see e.g., \cite{Yurke denker}). For an analysis of a detection without amplification see Refs. \cite{LesovikLoosen} and \cite{Gavish Levinson Imry}.}.
The main assumption used was that the conductance remains
approximately unchanged while the system is driven out of
equilibrium so that the latter remains impedance-matched to the
detector. This ensures that all the extra power emitted by the
shot-noise is detected. Claiming that such a measurement yields
$S(\omega)$  may seem to contradict the assumption Eq.\ref{S sym
def} however we shall now show that there is no inconsistency:
Also in the case of Eq.\ref{S sym def} the excess measurement
yields $S(\omega)$ and not $S_{sym}(\omega).$

By its definition the measured excess noise is
\begin{eqnarray}
\label{S exc def} S_{m,excess}(\omega)\equiv
S_{m}(\omega)-S_{m,eq}(\omega)\end{eqnarray}
 where $S_{m,eq}(\omega)$ is the noise measured in equilibrium.
Assuming Eq.\ref{S sym def}, we have
\begin{eqnarray}
\label{S exc 1} S_{m,excess}(\omega)= \frac{1}{2}
(S(\omega)+S(-\omega))-\frac{1}{2}
(S_{eq}(\omega)+S_{eq}(-\omega)).\end{eqnarray}
 This can be written as
\begin{eqnarray}
\label{S exc 2} S_{m,excess}(\omega)=S(\omega)-
S_{eq}(\omega)+\frac{1}{2} (S(-\omega)-S(\omega))-\frac{1}{2}
(S_{eq}(-\omega)-S_{eq}(\omega)).\end{eqnarray}
 Applying Eq.\ref{kubo noneq} we get
\begin{eqnarray}
\label{S exc 3} S_{m,excess}(\omega)=S(\omega)-
S_{eq}(\omega)+\hbar\omega (
\tilde{G}_d(\omega)-G_d(\omega)).\end{eqnarray}
 Since we assumed that the conductance remains the same in and out of equilibrium,
 the last term on the right  vanishes. and one is left with
\begin{eqnarray}
\label{S exc 4} S_{m,excess}(\omega)=S(\omega)-
S_{eq}(\omega).\end{eqnarray}
 Finally, since $S(\omega)$ is the emission spectrum \cite{Gavish Levinson Imry} and since in equilibrium at zero temperature there is no emission,
one has
\begin{eqnarray}
\label{S exc 5} S_{eq}(\omega)=0~~~~~~~~~~ k_BT=0\end{eqnarray}
 and thus
\begin{eqnarray}
\label{S exc 6} S_{m,excess}(\omega)=S(\omega)~~~~~~~~~~
k_BT=0\end{eqnarray}
 as asserted in \cite{Gavish Delft}.
Thus, also in the \emph{specific case} of Eq.\ref{S sym def}, the
excess noise measurement yields  the general results Eqs. \ref{S
exc 4} and \ref{S exc 6}. We emphasize that contrary to a common
view in the literature \cite{Landau Lifshitz}, Eq.\ref{S sym def}
is merely a specific case and not a general rule. For a concrete
example where Eq.\ref{S sym def} does not hold see
Ref.\cite{debloc}.

\section{Second application. Commutation relations for fermionic current components}
\label{sec 4} We now apply Eq.\ref{kubo noneq} in order to obtain
commutation relation for fermionic current components. Let us
define
\begin{eqnarray}
\label{Iw def}
I(\omega)=\frac{1}{\sqrt{2\pi}}\int_{-\infty}^{\infty} dt
e^{i\omega t}I(t).\end{eqnarray}
  $I(t)$ is hermitian and therefore
 \begin{eqnarray}
\label{Iw Iwdg} I^{\dag}(\omega)=I(-\omega).\end{eqnarray}
 In any stationary state one has:
\begin{eqnarray}
 \label{proof1} \langle I(\omega) \rangle= \frac{1}{\sqrt{2\pi}} \int_{-\infty}^{\infty} dt  e^{i\omega t }
\langle I(t)\rangle ~~~~~~~~~~~~~~~~~~~~~~~~~~~~~~ \nonumber\\
=\frac{1}{\sqrt{2\pi}} \int_{-\infty}^{\infty} dt  e^{i\omega t }
\sum_{i}P_ie^{iE_it}\langle i| I |i\rangle
e^{-iE_it}=\delta(\omega)\sqrt{2\pi} \langle I \rangle,
\end{eqnarray}
 where $P_i$ is the probability to be in the eigenstate $|i\rangle ,$
and
\begin{eqnarray}
 \label{proof2} \langle I(\omega) I(\omega')\rangle=~~~~~~~~~~~~~~~~~~~~~~~~~~~~~~~~~~~~~~~~~~~~~~~~~~~~~~~~~~~~~~~~~~~~~~ \nonumber\\ \frac{1}{2\pi} \int_{-\infty}^{\infty}\int_{-\infty}^{\infty} dt dt' e^{i\omega t +i\omega't'}
\sum_{i}P_ie^{iE_it}\langle i|I e^{-iH(t-t')}I|i\rangle
e^{-iE_it'}.\end{eqnarray}
 Defining $$\tau_+=\frac{1}{2} (t+t'),~~\tau_-=t'-t$$ $$t=\tau_++\frac{1}{2}\tau_-\\t'=\tau_+-\frac{1}{2}\tau_-,$$
one has
\begin{eqnarray}
 \label{proof3} \langle I(\omega) I(\omega')\rangle=~~~~~~~~~~~~~~~~~~~~~~~~~~~~~~~~~~~~~~~~~~~~~~~~~~~~~~~~~~~~~~~~~~~~~~ \nonumber\\ \frac{1}{2\pi} \int_{-\infty}^{\infty}\int_{-\infty}^{\infty} d\tau_- d\tau_+ \sum_iP_i
 \langle i|  I e^{iH\tau_-}I |i\rangle e^{-iE_i\tau_-}
 e^{i(\omega+\omega')\tau_+}
 e^{i(\omega-\omega')\frac{1}{2}\tau_-}\nonumber\\ = \int_{-\infty}^{\infty} d\tau_-
\langle   I e^{iH\tau_-}Ie^{-iH\tau_-}\rangle
\delta(\omega+\omega')
 e^{-\frac{i}{2}(\omega-\omega')\tau_-}\nonumber\\  = \delta(\omega+\omega') \int_{-\infty}^{\infty} d\tau_- e^{-i\omega\tau_-}
\langle  I(0) I(\tau_-)\rangle .\end{eqnarray}
  Thus,
 \begin{eqnarray}
\label{IIw}\langle I(\omega) I(\omega')\rangle  =
\delta(\omega+\omega') S(-\omega) .\end{eqnarray}

The averaged cosine and sine components of a current,
$\bar{I}_{cs}(\Omega)$ and $\bar{I}_{sn}(\Omega),$  are Hermitian
operators defined by
\begin{eqnarray}
 \label{cos and sine comp} \bar{I}_{cs}(\Omega)\equiv \bar{I}(\Omega)+\bar{I}^{\dag} (\Omega),~~ \bar{I}_{sn}(\Omega)\equiv -i(\bar{I}(\Omega)-\bar{I}^{\dag} (\Omega)),\end{eqnarray}
 where we defined averaging over a frequency bandwidth by
\begin{eqnarray}
\label{ave} \bar{X}(\Omega)=\ib d\omega X(\omega)~~~~~~~~~~ B:
[\Omega-\frac{1}{2} \Delta,\Omega+\frac{1}{2} \Delta]
.\end{eqnarray}
 In a current carried by a boson field, $\bar{I}_{cs}(\Omega)$ and $\bar{I}_{sn}(\Omega)$ form a canonical pair similar to $x$ and $p$
of an harmonic oscillator (and in that case $I(\Omega)$ is
analogous to the annihilation operator of an harmonic oscillator).
That is,
\begin{eqnarray}
\label{IcIs bos}
[\bar{I}_{cs}(\Omega),\bar{I}_{sn}(\Omega)]=if(\Omega)\end{eqnarray}
 where $f(\Omega)$ is a real c-number which may depend on $\Omega.$
An example for such a case is the current field in an ideal
transmission line \cite{LouisellTransLine},
\cite{YurkeDenkerTransLine}. Eq.\ref{IcIs bos} allows one to
derive uncertainty relations involving the current components
which have important consequences in the theory of quantum
amplification \cite{Yurke denker}\cite{caves1},\cite{caves2}.
However, this equation is generally \emph{not} valid for a current
carried by fermions, in which case the above commutator is in
general an operator.

To overcome this problem we shall use Eq.\ref{kubo noneq}.
Inserting Eq.\ref{Iw def} into \ref{S def}, integrating, averaging
over the band width $B,$ and making use of Eqs.\ref{Iw Iwdg} and
\ref{IIw} one gets
\begin{eqnarray}
\label{S ave} \sbr(\Omega)= \langle
\ibrd(\Omega)\bar{I}(\Omega)\rangle\nonumber\\ \sbr(-\Omega)=
\langle \bar{I}(\Omega)\ibrd(\Omega)\rangle.\end{eqnarray}
 Subtracting these two equations and making use of Eq.\ref{kubo noneq} result in
\begin{eqnarray}
\label{IcIs comm} \langle
[\bar{I}(\Omega),\ibrd(\Omega)]\rangle=2\hbar\Omega
\tilde{G}_d(\Omega)\Delta,\end{eqnarray}
 where for simplicity we assume that $\Delta$ is small enough so that $\tilde{G}_d(\Omega)$ remains constant in it.
From Eq.\ref{cos and sine comp} one sees
\begin{eqnarray}
\label{II IcIs com}
[\bar{I}_{cs}(\Omega),\bar{I}_{sn}(\Omega)]=2i[\bar{I}(\Omega),\ibrd(\Omega)]
.\end{eqnarray}
 Combining the last two equations we  finally get
\begin{eqnarray}
\label{IcIs comm} \langle
[\bar{I}_{cs}(\Omega),\bar{I}_{sn}(\Omega)]\rangle =4i\hbar\Omega
\tilde{G}_d(\Omega)\Delta.\end{eqnarray}
 Thus, we have transformed Eq.\ref{kubo noneq} into the form of commutation relations which is valid for any current in a stationary state, whether it
is carried by fermions or bosons. The usefulness of Eq.\ref{IcIs
comm} stems from the fact that in many cases, the conductance
$\tilde{G}_d(\Omega)$ is the same in a large set of stationary
states (as e.g. was the case in Sec.\ref{sec 3}) and therefore,
within such a set, $\bar{I}_{cs}(\Omega)$ and
$\bar{I}_{sn}(\Omega)$ posses  properties of an ordinary pair of
canonical variables.

\section{Summary and conclusions}
Kubo's fluctuation dissipation theorem holds also outside of
equilibrium, as long as the system is in a {\em stationary} state.
As a consequence, excess noise measurement of, for example,  the
symmetrized noise spectrum yields the nonsymmetrized one. Another
consequence is that although the commutator of the  two components
of fermionic currents is not in general a  purely imaginary
constant c-number (unlike for their bosonic counterparts), the
projection of their commutator onto all stationary states having
the same conductance, is such a c-number. This can be shown to
result in  Heisenberg constraints on the performance of quantum
transistors \cite{future}.

\vspace{0.2 cm}

We acknowledge essential discussions with Y. Levinson. This
project was partially supported by a Center of Excellence of the
Israel Science Foundation. It was also partially supported by the
German Federal Ministry of Education and Research (BMBF), within
the framework of the German Israeli Project Cooperation (DIP).


\begin{thebibliography}{99}
\bibitem{3D} In three dimensional notations, e.g. for a rectangular conductor, $L_x\times L_y\times L_z,$  with current flowing
along the $x$-direction, one should replace Eq.\ref{I def}, \ref{j def} and \ref{Gd def} by
$I=\frac{1}{L_x}\int_{L_x\times L_y\times L_z} dx dydz j(\vec{x}),~~$ $j(\vec{x})=\frac{e}{2m}\sum_i (P_{xi} \delta(\vec{x}-\vec{x}_i)+ H.c.)~~$
and \ $G_d(\omega)=Re \sigma (\omega)\frac{L_y L_z}{L_x}.$
\end{thebibliography}
\end{document}